\documentclass[aps,prb,twocolumn,groupedaddress,floatfix]{revtex4}

\usepackage[dvips]{graphicx}
\usepackage{verbatim}

\def\it{\textit}
\def\bf{\textbf}
\def\Tc{$T_{c}$}

\def\sig1{\sigma_{1}}

\def\TCO{Ti$_{1-x}$Co$_{x}$O$_{2-\delta}$}
\def\TO{TiO$_2$}
\def\LAO{LaAlO$_3$}

\def\cm{cm$^{-1}$}
\def\T{\begin{mathcal}T\end{mathcal}}
\def\R{\begin{mathcal}R\end{mathcal}}
\def\O2{O$_{2}$}
\def\etal{\textit{et al}.}

\begin{document}

%\preprint{}

% ********************** Title **************************************
\title{Optical band edge shift of anatase \TCO}


\author{J. R. Simpson}
\email[]{simpson@physics.umd.edu}
\author{H. D. Drew}
\affiliation{\it{Department of Physics, University of Maryland,
College Park, Maryland 20742-4111}}
\author{S. R. Shinde}
\author{R. J. Choudhary}
\author{S. B. Ogale}
\author{T. Venkatesan}
\affiliation{\it{Center for Superconductivity Research, Department of Physics, University of Maryland, College Park, Maryland 20742-4111}}
%\homepage[]{http://www.glue.umd.edu/~simpson/}
%\thanks{}


% date for con-mat archive 
\date{August 27, 2003}
%\date{\today}

% *********************** Abstract **********************************
\begin{abstract}

We report on the optical properties of magnetic cobalt-doped
anatase phase titanium dioxide \TCO\ films for low doping
concentrations, $0 \le x \le 0.02$, in the spectral range $0.2 \le
\hbar \, \omega < 5$~eV.  For well oxygenated films ($\delta \ll
1$) the optical conductivity is characterized by an absence of
optical absorption below an onset of interband transitions at
3.6~eV and a blue shift of the optical band edge with increasing Co concentration.
The absence of below band gap absorption is inconsistent
with theoretical models which contain midgap magnetic impurity bands and suggests that strong on-site Coulomb interactions shift the O-band to Co-level optical transitions to energies above the gap.
\end{abstract}

%\keywords{}

\maketitle

% ********************** Introduction *******************************
%\section{\label{intro}Introduction}

Dilute magnetic semiconductors (DMS) offer a possible system to
realize control of the charge transport by using the spin degrees
of freedom or "spintronics".\cite{ohno98}  DMS consist of magnetic
impurities doped in a semiconducting host (e.g., Mn-doped GaAs).
Such materials undergo a ferromagnetic phase transition below the
Curie temperature, \Tc\ which is typically \mbox{$T_{c} \approx
100$~K}.

Therefore the recent discovery of ferromagnetism with \mbox{$T_{c}
>300$~K} in cobalt-doped \TO\ has generated considerable interest
in this system and similar dilute magnetic
oxides.\cite{matsumoto01}  High-temperature magnetization
measurements using vibrating sample magnetometry\cite{sanjay} find
a $T_{c} \gtrsim 1180$~K for $x=0.07$ \TCO, nearly that of bulk Co
($T_{c} = 1404$~K).  Such a large $T_c$, suggests that cobalt
appears in clusters rather than substitutionally and that the
resulting magnetism is due to clustered Co rather than a new dilute
magnetic oxide.  Indeed several groups report direct observation
of cobalt clusters from transmission electron
microscopy.\cite{chambers01,sanjay,park02}  In a careful study of doping
dependence, Shinde \etal\cite{sanjay} find a limited solubility of
cobalt in \TCO\ above a concentration of $x \sim 0.02$ with Co
clustering beginning thereafter.  Low-doped systems, $x \le 0.02$,
exhibit a $T_{c} \approx 700$~K and show no evidence of Co
clusters.  The existence of such a high \Tc\ for a dilute magnetic
system showing no Co clustering remains puzzling. Thus, further
measurements to elucidate the electronic structure and resolve the
nature of the magnetism are warranted.

In this paper, we present measurements of the optical absorption
of well-characterized thin films of anatase \TCO\ for low Co
concentrations.  Interband absorption above the band gap at 3.6~eV
dominates the optical spectra.  We discuss the implications of
these results related to band structure calculations and compare
our measured band edge shift to other optical studies.

% ******************* Experiment ************************************
%\section{\label{experiment}Experiment}

Thin film samples of \TCO\ with $x=0, 0.01$, and $0.02$ were grown
on SrLaGaO$_4$ (SLGO) substrates using pulsed laser deposition.
Additionally, a pure \TO\ film was grown on LaAlO$_3$ (LAO) for
comparison. Films were deposited to $\sim 1500$~\AA\ thickness
with an oxygen partial pressure of 10$^{-5}$~Torr corresponding to
($\delta \ll 1$). 4-probe dc resistance measurements exhibit
insulating behavior with room temperature resistivity
\mbox{$\rho_{295\text{K}} \gtrsim 10\ \Omega\,$cm}.  X-ray diffraction
(XRD) measurements\cite{sanjay} of both pure and doped \TO\ films
show peaks corresponding to those observed in bulk anatase \TO.
Values of the in-plane and out-of-plane ($d_{004}$) lattice
constants obtained from XRD are discussed later in the paper.

Room temperature transmission $\T(\omega)$ and reflection
$\R(\omega)$ measurements of near-normal incidence light at
frequencies from 0.25 to 5~eV are performed using a
Fourier-transform spectrometer.\cite{quijada98, simpson99} By
numerically inverting the Fresnel formulas\cite{heavens91} for
$\T$ and $\R$, we obtain the complex index of refraction
$\tilde{n}=n+i\,k$ without the need for Kramers-Kronig
analysis.\cite{quijada98, simpson99} From $\tilde{n}(\omega)$, we
may derive other optical constants, e.g. the optical absorption
$\alpha(\omega)$ or the complex optical conductivity
$\tilde{\sigma}(\omega)$.

\begin{figure}[ht]
\begin{center}\leavevmode
\includegraphics{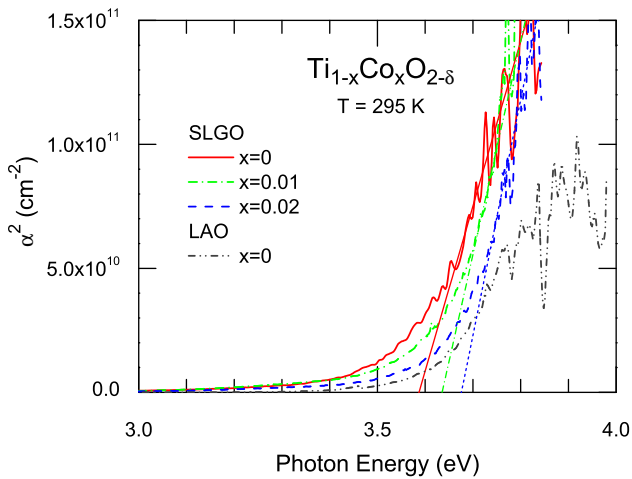}
\caption{Frequency dependence of the square of the absorption
coefficient $\alpha$ at room temperature for Co concentrations
$x=0$, 0.01, and 0.02.  Straight lines represent linear fits.  An
undoped \TO\ sample grown on a \LAO\ substrate is shown for
comparison.} \label{fig;alphasq}
\end{center}
\end{figure}

Historically, the spectral dependence of the band edge is
characterized using the absorption coefficient, $\alpha = 4
\pi\,\omega\,k$, where $\omega$ is the frequency in \cm\ and $k$
is the extinction coefficient ($k=$ Im\{$\tilde{n}$\}).  At photon
energies above the band gap $E_g$, $\alpha \propto (\hbar\omega -
E_g)^{1/2}$ for a direct gap while $\alpha \propto (\hbar \omega -
E_g)^{2}$ for an indirect gap.\cite{yu-cardona}  Band structure
calculations\cite{asahi00} predict a direct gap at energies just
lower than the onset of indirect transitions.  In a detailed study
of the absorption edge of single crystal anatase \TO, Tang
\etal\cite{tang95} report a band edge with $E_g = 3.420$~eV and
tentatively assigns the transition to a direct gap.  Consistent
with the behavior of a direct gap, we plot $\alpha^2$ versus
frequency in Fig.~\ref{fig;alphasq}.
\noindent Linear fits to the high-frequency part above the band
edge onset are extrapolated to zero absorption, giving the direct
band gap energy $E_g$.  Pure \TO\ exhibits $E_g = 3.6$~eV,
slightly larger than the value obtained by Tang \etal\cite{tang95}
With increased Co concentration $x$, the band edge shifts to
higher frequencies, showing a maximum shift of 100~meV for
$x=0.02$.  Apart from a band edge tail, there appears no evidence
for strong absorption at frequencies below the gap.  Additionally,
we compared our results to linear fits of $\sqrt{\alpha}$
consistent with an indirect gap and found that while the indirect
band edge is approximately 0.4~eV lower in energy, the shift of
$E_g$ with doping remains unaffected.  The exact nature of the gap
(direct or indirect) remains uncertain.

\begin{figure}[hbt]
\begin{center}\leavevmode
\includegraphics{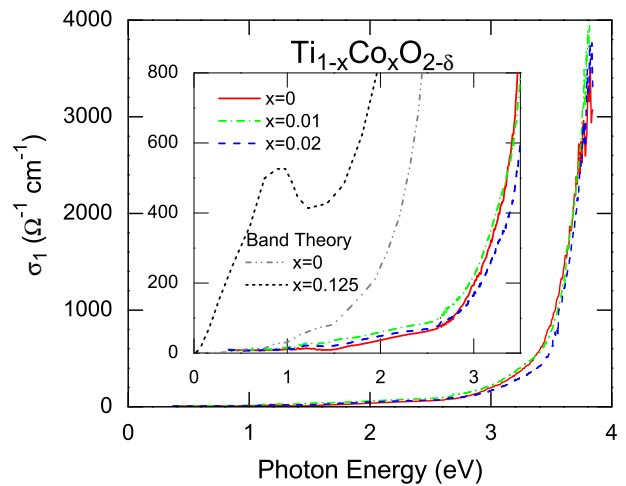}
\caption{Frequency dependence of the real part of the optical
conductivity $\sigma_1$ at room temperature for Co concentrations
$x=0$, 0.01, and 0.02. Inset expands the region just below the
band edge.  Band theory calculations from Yang \etal\cite{yang03}
are shown for $x=0$ and $x=0.125$.} \label{fig;sigma1}
\end{center}
\end{figure}

While $\alpha$ well characterizes the band edge onset, the optical
conductivity better suites to compare experimental results to
predictions of band theory.  In particular, the real part of the
optical conductivity is given by $\sig1(\omega) \propto
2\,n\,k\,\omega$, where $n$ and $k$ are the real and imaginary
part of the complex index of refraction respectively and $\omega$
is the frequency.  Figure~\ref{fig;sigma1} shows the frequency
dependence of $\sig1$ at room temperature for several Co
concentrations. Throughout the mid-infrared (mid-IR) to visible
frequency range (0.25 to 3~eV), $\sig1$ remains essentially zero,
consistent with the negligible dc conductivity ($\sigma_{dc}
\lesssim 0.1\ \Omega^{-1}\,$\cm). At frequencies larger than 3~eV,
$\sig1$ increases rapidly corresponding to the increase in
absorption as seen in Fig.~\ref{fig;alphasq}. To further elucidate
the minimal midgap conductivity, we expand the scale in the inset
of Fig~\ref{fig;sigma1}.  Although no strong conductivity in the
spectral range $1 \le \hbar\omega \le 3$ is observed, $\sig1$
increases slightly near the edge, although non-monotonically, with
the addition of Co.  Such an increase may result from the Co
levels in the gap or disorder effects on the Urbach
tail\cite{tang95} of the fundamental absorption edge.
%

% ******************* Analysis **************************************
%\section{\label{analysis}Analysis}

\begin{figure}[hbt]
\begin{center}\leavevmode
\includegraphics{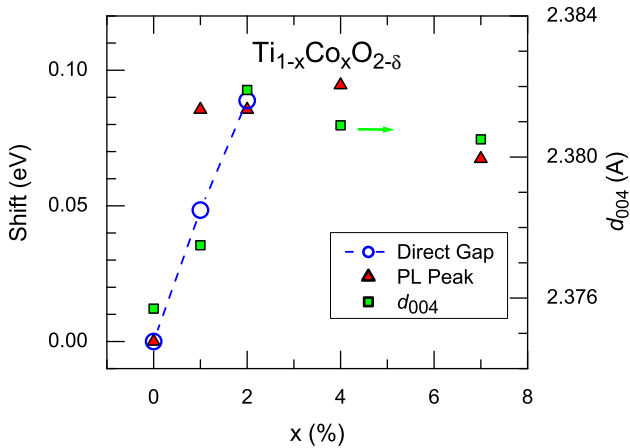}
\caption{Shifts with cobalt doping of the direct band edge
(circles with a dashed line as a guide for the eye) and
photoluminescence peak from Guha \etal\cite{guha03} (triangles).
For comparison the out of plane lattice constant,\cite{sanjay}
$d_{004}$, for films grown on LaAlO$_3$ is plotted on a separate
y-axis (squares).} \label{fig;shifts}
\end{center}
\end{figure}

% PL DATA
It is interesting to compare the optical conductivity and band
edge shifts with photoluminescence (PL) spectra. PL
studies\cite{guha03,tang93,tang94b} find a broad peak centered
around 2.3~eV for anatase \TO.  The peak in the PL spectrum is
Stokes shifted 1.3~eV lower than the onset of interband
transitions in the optical conductivity and the peak in the
photoluminescence excitation spectrum.\cite{tang94b}  Such a Stokes
shift is consistent with that of a self-trapped exciton where the
exciton loses energy nonradiatively to the lattice. With the
addition of Co, the peak of the PL spectrum blue
shifts,\cite{guha03} similar to the shift observed in the band
edge. Doping dependent shifts in the band edge and the PL peak are
plotted together in Fig.~\ref{fig;shifts} for comparison. The direct band
gap energy increases monotonically with $x$, while the PL peak
increases rather abruptly upon the addition of Co and saturates
above $x=0.02$.  The saturation of the shift at higher doping
concentrations is consistent with the limited solubility of Co in
\TO\ as reported earlier.\cite{sanjay}
%

% PRESSURE-DEPENDENT OPTICS
Similar shifts of the band edge as observed here with Co doping
are reported in a pressure-dependent optical study of single x-tal
anatase \TO.  Sekiya \etal\cite{sekiya01} observe a blue shift of
the band edge upon the application of hydrostatic pressure.  For
an applied pressure of 3.9~GPa, the edge shifts to higher energy
by $\sim 50$~meV.  We estimate the change in the \TO\ lattice
resulting from the hydrostatic pressure by introducing the bulk
modulus, $B = - V \Delta P / \Delta V$, where $V$ is the volume,
$\Delta V$ is the change in volume, and $P$ is the applied
pressure.  Taking the bulk modulus for anatase \TO,
$B=180$~GPa,\cite{arlt00} the applied hydrostatic pressure of
3.9~GPa introduces a volume decrease of $~2$\%.

% SUBSTRATE STRAIN
Given the blue shift of the band edge with increasing pressure or
equivalently decreasing lattice size, we consider how substrate
strain or the addition of cobalt may affect the lattice.  Firstly
the issue of substrate strain, the thin films studied suffer
strain due to lattice mismatch with the substrates.  For epitaxial
films the initial layers grow coherently with the substrate but
defects gradually relax the induced strain allowing the films to
grow more like bulk.  Values of the in-plane lattice constant,
$a$, for bulk \TO, LAO, and SLGO are listed in
Table~\ref{tab;lattice}.
\begin{table}[htb]
\caption{Lattice parameters.} \label{tab;lattice}
\begin{ruledtabular}
\begin{tabular*}{3.4in}{l @{\extracolsep\fill} r@{\extracolsep{0pt}}l @{\extracolsep\fill}  r@{\extracolsep{0pt}}l}
  & \multicolumn{2}{c}{$a$ (\AA)} & \multicolumn{2}{c}{$d_{004}$ (\AA)}  \\ \hline
\TO\ Bulk\footnote{Ref.~\onlinecite{arlt00}} & 3.&7851    &  2.&3780\\
LaAlO$_3$ substrate   & 3.&79  &  & \\
SrLaGaO$_4$ substrate & 3.&84  &  & \\
\TO\ film on LaAlO$_3$   & & & 2.&376\,\footnote{X-ray diffraction measurement.} \\
\TO\ film on SrLaGaO$_4$ & & & 2.&367\,\footnotemark[2] \\
\end{tabular*}
\end{ruledtabular}
\end{table}
Both LAO and SLGO substrates have a larger in-plane lattice
parameter than bulk \TO, introducing a tensile strain in the films
of approximately 0.12\% and 1.5\%, respectively.  The tensile
stress tends to expand the \TO\ lattice in the plane.  For
materials with a typical Poisson ratio, an expansion in the plane
results in a reduction of the out-of-plane lattice constant,
$d_{004}$. Indeed XRD measurements of the \TO\ films reveal such a
decrease of $d_{004}$, as shown in Table~\ref{tab;lattice}.  The
film on SLGO, with the larger in-plane tensile strain, exhibits a
larger reduction in $d_{004}$ relative to bulk \TO\ ($\sim
0.45\%$) compared with the film on LAO ($\sim 0.08\%$). To explore
the effects of lattice strain on the band edge, the absorption of
a \TO\ film grown on LAO is shown in Fig.~\ref{fig;alphasq}.  The
band edge of the film on LAO with the smaller in-plane lattice is
blue shifted ($\gtrsim 100$~meV) relative to the \TO\ film grown
on SLGO, consistent with the blue shift resulting from the
application of hydrostatic pressure discussed above.

% LATTICE EXPANSION W/ CO ADDITION
To address the effect of cobalt substitution on the shift of the
band edge, we examine the change in size of the lattice with
doping.  Cobalt appears in the doped \TO\ system in the +2 formal
oxidation state as determined from x-ray absorption
spectroscopy.\cite{chambers01}  The atomic radii of Co$^{2+}$ and
Ti$^{4+}$ are 0.82~\AA\ and 0.69~\AA, respectively.\cite{emsley98}  Substitution of the larger Co$^{2+}$ for
Ti$^{4+}$ should expand the lattice.  XRD
measurements\cite{sanjay} of films grown on LAO (plotted in
Fig.~\ref{fig;shifts}) show $d_{004}$ increases with $x$,
saturating at about $x=0.02$.  The increase of $d_{004}$ supports
the prediction of an increase in the size of the lattice with Co
doping.  For the films grown on SLGO, $d_{004}$ remains relatively
constant with Co, $d_{004}=2.3674 \pm 0.0002$~\AA.  An increasing
(films on LAO) or relatively constant (films on SLGO) lattice size
with $x$ should result in either a red shift or no shift of the
band edge.  Therefore, the observed blue shift with Co doping
cannot be ascribed simply to a change of the lattice size.

% BAND STRUCTURE AND ELECTRONIC TRANSITIONS
In order to understand the observed blue shift of the band edge,
we compare our conductivity to theoretical predictions of the
electronic structure of both pure and Co-doped \TO.  Band
structure calculations\cite{mo95,park02} indicate the valence band
derives primarily from oxygen p-levels, the conduction band
derives from the Ti d-levels, and that the crystal-field split Co
d-levels fall within the energy gap.\cite{park02}  These midgap
states would lead to below band gap optical absorption in a
non-interacting electron picture of optical transitions. Using a
first-principles density-functional approach, Yang
\etal\cite{yang03} calculates the dielectric function for Co-doped
\TO.  The resulting $\sigma_1$  for $x=0.125$ and pure \TO\ are
shown in the inset of Fig.~\ref{fig;sigma1} for comparison.

The predicted increase in conductivity below the band gap results
from transitions to cobalt levels.  We may estimate the effective
Co number density using optical sum rules.
\begin {equation} \label{eq;sw}
S(\omega) = \frac{2}{\pi} \int_{0}^{\omega} \sigma_1(\omega') \,
d\omega' = \frac{N_{eff}\,e^{2}}{m} \ ,
\end{equation}
\noindent where $N_{eff}$ is the effective carrier density, which
in general will be somewhat less than the cobalt number density
$N$, $e$ is the electron charge, and $m$ is the electron mass.
The cobalt number density $N$ as a function of $x$ is given by $N
= \frac{f x}{V}$, where $f=4$ is the number of Ti per unit cell
and $V=136.85$~\AA$^{3}$ is the unit cell volume.  Estimating
$N_{eff}$ from the predicted conductivity using Eq.~\ref{eq;sw}
and comparing to $N$, we find that approximately 0.75 of the total
Co spectral weight appears in the predicted midgap absorption
feature.  In estimating the Co spectral weight from the
experimental data we take the difference in $\sig1$ due to doping
to be a constant $\sim 10 \; \Omega^{-1}\,$\cm\ (corresponding to
the error in our measurement) over the frequency range from 1 to
3~eV.  Substituting into Eq.~\ref{eq;sw}, we find the experimental
upper bound for $N_{eff} \sim 10^{20}$~cm$^{-3}$. Comparing this
to the number density $N$ we find the \it{upper} bound on the
observed density of Co is roughly 0.15 times the expected total.

The absence of below gap optical excitations in the measured
conductance may be understood either as evidence that the band
calculations are not capturing the electronic structure of this
material under the assumed charge state of the cobalt or that the
on-site Coulomb energy, $U$, for adding another electron to the Co
ion is large.  In the second scenario the experiment implies $U
\gtrsim 3$~eV.

\begin{figure}[ht]
\begin{center}\leavevmode
\includegraphics[clip]{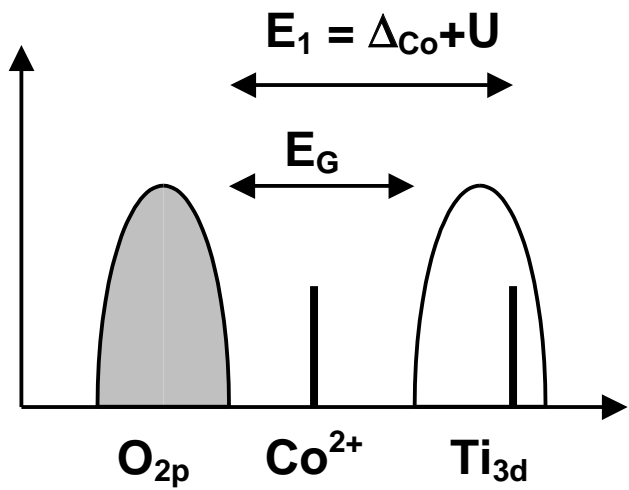}
\caption{Schematic energy level diagram.} \label{fig;levels}
\end{center}
\end{figure}

Noting the absence of spectral weight associated with cobalt
levels in the gap combined with the blue shift of the band edge,
we examine the possible strong interaction effects on the optical
transitions involving the Co ion.  First we reject interpretations
that consider the alloy within a rigid band picture. In this case
a shift in the band edge results from uniform shifts of the
conduction band due to the average Ti/Co potentials.  Since the
atomic potentials for Co are larger that those of Ti, the Ti/Co
band would be lower than the pure Ti bands in \TO\ resulting in a
red shift of the band edge, contrary to observation.  Indeed the rigid
band approach is more appropriate for delocalized states.  For the
transition metal ions in \TCO\ a localized picture is more
appropriate.  Therefore, we discuss the processes operating on the
optical transitions involving the Co levels within a localized
picture. A schematic view of the band structure is shown in
Fig.~\ref{fig;levels}.
In pure \TO, the band edge $E_g$ consists of the energy difference
between the filled O~p-levels and the empty Ti~d-levels with $E_g
\approx 3.6$~eV (as discussed above).  With the addition of
cobalt, charge transfer transitions from the O~p-levels to the
empty localized Co$^{2+}$ d-levels become possible.  Allowed
transitions from occupied Co levels to the Ti levels should be
weaker since they involve a virtual transition through the
O~p-levels. The energy $E_1$ of the oxygen to cobalt transition is
the sum of the charge transfer energy $\Delta_{\text{Co}}$ plus
the on-site Coulomb energy $U$; $E_1= \Delta_{\text{Co}}+U$.  The
observed absence of below band gap absorption indicates that $E_1$
is greater than the band gap in the alloy $E_g'$.  This is
reasonable since the $U$ is estimated to be $\sim 3$~eV and the
empty Co levels are $\sim 2$~eV above the oxygen
band.\cite{park02}

We now turn our attention to the observed band edge shift, $\Delta
E_g = \xi\,x$, where $\xi=5$~eV.  If the O and Ti band edges are
not affected by the substitution of Co, this shift would be
understood in terms of the reduction of the interband oscillator
strength upon Ti dilution by Co and the extra absorption at $E_1$.
However, this scenario leads not to a shift in the band edge but
essentially to a change in the slope of $\alpha^2$, contrary to
observation. Therefore we conclude that O and Ti bands separate
upon Co substitution.  The large rate of separation ($\xi=5$~eV)
implies strong level repulsion that might occur for interstitial
incorporation of the Co. This large band edge shift is especially
interesting because it implies strong interactions which are also
required to provide the large exchange interaction and associated
high ferromagnetic $T_c$ observed in this material.
%
%
% ****************** Conclusions ************************************
%\section{\label{conclusions}Conclusions}

In conclusion, optical measurements have revealed a shift of the
band edge with Co doping and an absence of midgap absorption in
anatase \TCO.  The gap in the optical conductivity also implies
strong Coulomb interaction effects on the optical processes
involving the Co ions.
%

% ****************** Acknowledgements *******************************

\begin{acknowledgments}
We wish to thank A. J. Millis, S. Das Sarma, and  G. A. Sawatzky
for valuable discussions and Y. Zhao for help with experimentation.  This work supported in part by
NSF-MRSEC Grant No. DMR-00-80008 and DARPA (S.D.S.).
\end{acknowledgments}

% ****************** Bibliography ***********************************

\bibliography{cto}

% Create the reference section old style (with \bibitem).
\begin{comment}

\end{comment}

\end{document}